# Human Mars Exploration and Expedition Challenges

Malaya Kumar Biswal M[*] and Ramesh Naidu Annavarapu[†]

*Department of Physics,
School of Physical Chemical and Applied Sciences,
Pondicherry University, Kalapet, Puducherry, India – 605 014*

Mars is the next frontier for the space explorers to demonstrate the extent of human presence in space beyond low-earth orbit. Both government and private space industries have been fascinated by Mars quest to attempt a crewed expedition to the red planet. The journey to Mars is vastly challenging as it endowed with numerous challenges from the inception of the mission engage to the mission achievement. Therefore, it is substantial to overcome those challenges for a reliable mission. Hence we have studied and emphasized the comprehensive challenges under the categorization of terrestrial, Earth-bound, interplanetary, Mars-bound, and planetary surface challenges. These challenges are suspected to encounter by the astronauts and mission planners throughout the mission timeline. Our research is different from other studies as it reports complete challenges and their implications on the way to human exploration of Mars.

## I. Nomenclature

| | | |
|---|---|---|
| *AMASE* | = | Arctic Mars Analog Svalbard Expedition |
| *AU* | = | Astronomical Unit |
| *CAVES* | = | Cooperative Adventure for Valuing and Exercising human behaviour and performance skills |
| *CFM* | = | Cryogenic Fuel Management |
| *CSA* | = | Canadian Space Agency |
| *D-MARS* | = | Desert Mars Analog Ramon Station |
| *D-RATS* | = | Desert Research and Technology Studies |
| *DSN* | = | Deep Space Network |
| *EDL* | = | Entry, Descent, and Landing |
| *EMU* | = | Extra-Vehicular Maneuvering Unit |
| *ESA* | = | European Space Agency |
| *Euro-Mars* | = | European Mars Analogue Research Station |
| *EVA* | = | Extra-Vehicular Activity |
| *FMARS* | = | Flashline Mars Arctic Research Station |
| *GCR* | = | Galactic Cosmic Radiation |
| *GNC* | = | Guide Navigation and Control |
| *HI-SEAS* | = | Hawaii Space Exploration Analog and Simulation |
| *HLLV* | = | Heavy Lift Launch Vehicle |
| *HMO* | = | High Mars Orbit |
| *HMP* | = | Haughton Mars Project |
| *IBP* | = | Institute of Biomedical Problems |
| *IMLEO* | = | Initial Mass in Low Earth Orbit |
| *ISRO* | = | Indian Space Research Organization |
| *ISS* | = | International Space Station |
| *LEO* | = | Low Earth Orbit |
| *MARIE* | = | Mars Radiation Environment Experiment |
| *Mars-Oz* | = | Australia Mars Analogue Research Station |

[*] Graduate Researcher, Department of Physics, Pondicherry University, India; malaykumar1997@gmail.com, mkumar97.res@pondiuni.edu.in, Student Member of Indian Science Congress Association, Student Member AIAA
[†] Associate Professor, Department of Physics, Pondicherry University, India; rameshnaidu.phy@pondiuni.edu.in. Non-Member AIAA.

| | | |
|---|---|---|
| *MAVEN* | = | Mars Atmosphere and Volatile Evolution |
| *MDRS* | = | Mars Desert Research Station |
| *MGS* | = | Mars Global Surveyor |
| *MLV* | = | Modified Launch Vehicle |
| *MOI* | = | Mars Orbital Insertion |
| *MOLA* | = | Mars Orbiter Laser Altimeter |
| *MOM* | = | Mars Orbiter Mission |
| *MRO* | = | Mars Reconnaissance Orbiter |
| *MSO* | = | Mars-Centered Solar Orbit |
| *MTO* | = | Mars Telecommunication Orbiter |
| *NASA* | = | National Aeronautics and Space Administration |
| *NEEMO* | = | NASA Extreme Environment Mission Operation |
| *NEP* | = | Nuclear Electric Propulsion |
| *NIMP* | = | Nuclear Rocket using Indigenous Martian Fuel |
| *NTG* | = | Nuclear Thermoelectric Generator |
| *NTR* | = | Nuclear Thermal Rocket |
| *PLRP* | = | Pavillion Lake Research Project |
| *RLV* | = | Reusable Launch Vehicle |
| *ROSCOSMOS* | = | Russian State Corporation for Space Activities |
| *RTG* | = | Radioisotope Thermoelectric Generator |
| *SCR* | = | Solar Cosmic Rays |
| *SEP* | = | Solar Electric Propulsion |
| *SLS* | = | Space Launch System |
| *SPE* | = | Solar Particle Events |
| TD | = | Technology Demonstration |
| *TGO* | = | Trace Gas Orbiter |

## II. Introduction

Space exploration has diverse challenges [1, 2] and the exploration of Mars possess numerous challenges whereas the Mars exploration has been a source of inspiration to global space firms for decades. The inspiration for Mars began from the conceptual and architectural designs first put forwarded by Wernher von Braun [3]. To date, we have extensively demonstrated numerous space technologies efficient for executing human class missions to Mars and beyond. Similarly, we have successfully explored the red planet with numerous autonomous spacecrafts and planetary probes [4-6]. So, from the perspective of sustainable and efficient human-crewed mission, it is substantial to consider the natural challenges caused by the phenomena of the space environment and technological challenges ensued from the artificial technologies. Hence, we have studied and emphasized every possible challenge that the crew may experience during their interplanetary spaceflight from the Earth to the Mars except the Entry, Descent, and Landing challenges, as it was technically reviewed by R.D. Braun [7]. Further, overall challenges were stratified into terrestrial, earthbound, interplanetary, Mars-bound, and planetary challenges under the simplified categorization of terrestrial, interplanetary, and planetary challenges outlined in Fig 1.

## III. Terrestrial Challenges

### A. Consideration of feasible future technology

From the perspective of efficient human-crewed Mars mission, it is significant to consider simple, robust, feasible, and affordable future technologies. It is not obvious to suspect our assumptions towards extreme or massive technology but to have substantial technology that can be engineered with our existing pieces of machinery. (For example: if we propose to use SEP (Solar Electric Propulsion) for the space transportation system, we might be having hardship in fabricating massive solar panels, enfolding together to place in LEO with the launch vehicle and deploying in interplanetary space. Further, the challenge for this type of propulsion system is the availability of solar irradiance for powered propulsion that gradually decreases as we move far apart from the sun with a variance of 1360 W/m$^2$ (at Earth) to 590 W/m$^2$ (at Mars). Alternatively, Electric Propulsion, NTR, NEP, and NIMP for transit to Mars can be exploited with the technology of controlled propulsion and management. It minimizes the risk of mass constraints [8-11].

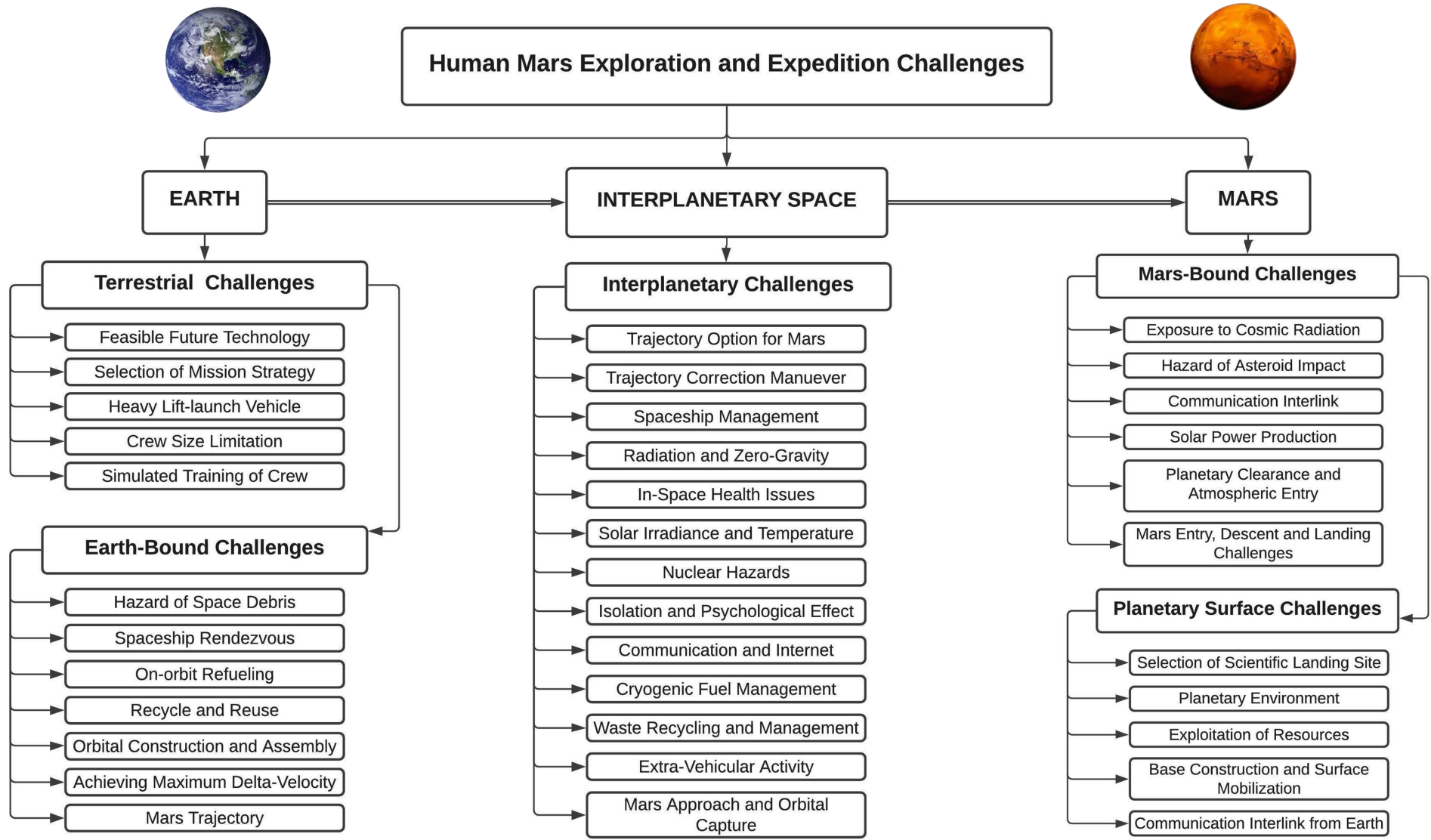

**Fig 1 Outline Map for Human Mars Exploration and Expedition Challenges**

Similarly, other than powerful propulsion systems, we need to develop and demonstrate advanced inflatable heat shield to ground large scale masses, power spacesuits to keep the astronauts safe against hostile Mars environment, the concept of sub-surface habitats to maintain thermal stability and the habitat on wheels to enable surface mobilization, uninterrupted power generation employing radioisotope thermoelectric generators or fission power systems, and advanced laser-guided communication system to stay tethered from Earth to receive more information and to stay updated about the mission strategies.

**B. Mission Design and Architecture Selection**

Mars enthusiasts and scientists have proposed more than 70 human Mars mission's architectures since 1952. Analyzing through hundreds of pages, comparing with feasible technologies, and executing the mission plan with the right budget seems complicated. Because each and every proposal have their desired goal. So it is preferable to sort out simple, robust with economic standard and feasible plans. Hence we have selected (12) architectural strategy for budget comparison and reliability shown in Fig 2.

Among these strategies, SpaceX's Mars expedition, Mars One, NASA's Design reference mission has the potential technology to attempt for a human-crewed mission to Mars within the time frame of 2040 and 2060s. Further, based on the current state of matured technology, these strategies were found to be feasible and inexpensive as compared to other mission plans.

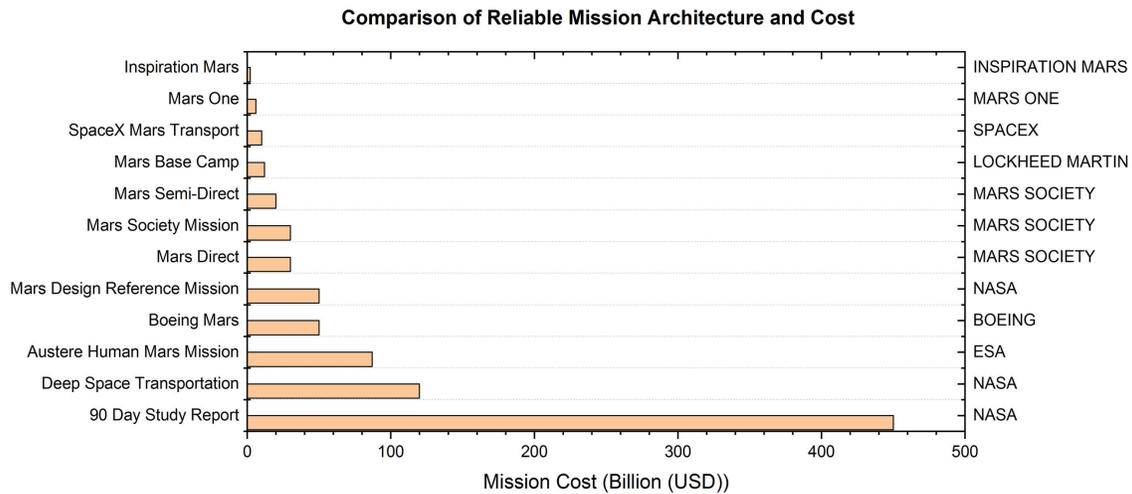

**Fig 2 Mission Architecture and Estimated Cost (in Billion USD).**

**C. Consideration of the Heavy Lift Launch Vehicle**

The primary progression step towards the Mars expedition is the launch of heavy mass (cargoes, crews, and necessities) to LEO as part of IMLEO. Delivering of large mass facilitates the reliability of mission thereby expanding the number of cargoes required for the crews and planetary exploration, size of the crew, and possible payloads for spaceship rendezvous in low earth orbit (LEO). But the technology of launch vehicle to deliver large scale mass has made a significant impact to execute a mission. Numerous architectures on launch vehicle enhancement were proposed like Soviet Union's N1 Rocket, Aelita, United States Saturn MLV, Comet HLLV, Ares, Sea Dragon, and SpaceX's Interplanetary Transportation System. These launchers may have the potential to deliver huge IMLEO mass with an extent from 100-1400 metric tons [12, 13]. But the task of construction, launch, testing, and validation may cost expensive and infeasible. Hence technologically feasible launchers such as Saturn V, Ares V, SLS, Falcon Heavy, Long March, Starship, New Glenn are considerable for manned mission [14-17]. It is because some are under development and some are technologically proven in past decades. Moreover, if these launchers come accessible, they can lift a mass variant from 40-200 metric tons to LEO and may pave a way for the affordable human class Mars mission. Launchers and their respective payload mass to LEO and launch cost comparison are shown in Fig 3.

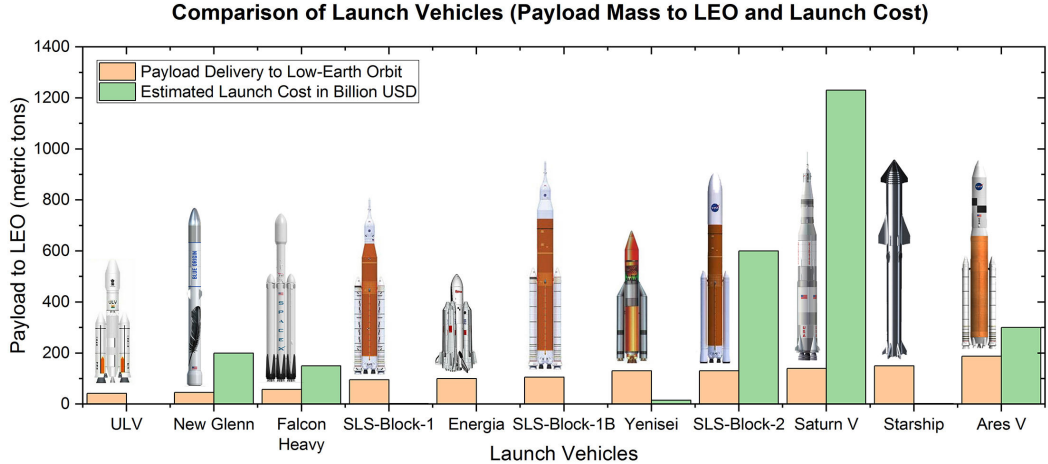

**Fig 3 Launch Vehicles and their payload capacity to LEO.**

### D. Crew Size Limitation

The most crucial part of the Human Mars Expedition is the commission of crew and crew size that determines the extent of mission accomplishment. The size of the crew defines the quantity of cargoes desired over the mission period that affecting the initial mass (IMLEO) allocated for launch. Comparably a limited number of crew influences the state of psychological fitness and the ability to accomplish the scientific goal. So it is substantial to sort out each crew with multitudinous skills (For example a crew capable of accomplishing his/her assigned work along with the skill of medical first aid, managing spaceship, rectifying the damaged instruments and miscellaneous). It is always desirable to have stability between the crew and the commodity. Hence we interpret that the limit of crew size from 4 to 6 seems to be ideal for mission robustness [18].

### E. Simulated Training of Crew

Before executing human class Mars mission in a challenging space environment, the astronaut (future Martians) undergoes a various and complex process of sub-orbital training to validate the rate of mission accomplishment during their stay in mission. For this intention, various space and Mars firms have effectuated human analog missions to simulate various environmental aspects. Some of the human Mars analog training centers with their locations and objectives are shown in table 1.

It is worth and satisfactory to have testing, validation, and training in a planetary simulated environment. But the results are incomparable to the actual space environment. Because some results may go erroneous. In a long-duration space mission, the first and foremost priority must be given to the crew safety and surviving requirements. For every emergency, there must be a back-up plan and mission abort option that can be proceeded. We cannot put life at risk on a voyage of the search for life. It is not that the mission can be accomplished without crew, instead, it is with the crew. It is significant to make avail every crew necessity at a remote distance [19-22].

**Table 1 – List of Mars Analog Research Stations**

| Name | Location | Agency | Objectives |
|---|---|---|---|
| D-RATS | Arizona Desert | NASA | Hardware and Rover Testing, Sample Collection |
| PLRP | Pavillion Lake | NASA/CSA | Test Deep Space Mission Concepts |
| NEEMO | Wilmington | NASA | Simulate and test low-gravity experiments |
| AMASE | Norway | NASA | Testing of Hardware and Instruments, Biological Experiment. |
| FMARS | Devon Island | Mars Society | Martian Surface Simulation |
| MDRS | Southern Utah | Mars Society | Hardware testing and conceptual designs |
| HMP | Devon Island | Mars Society | Test new space technologies |
| Mars 500 | Moscow | RAS/IBP | Psychological Implications of long spaceflight |
| ESA-CAVES | Europe | ESA | Train astronauts in the cave environment |
| D-MARS | Negev Desert | Israel/Austria | Robotic experiments |
| HI-SEAS | Hawaii | NASA | Research in food, crew and dynamics, behavior, roles and performance |
| Euro-Mars | Iceland | Mars Society | Test habitats and space tools |
| Mars-Oz | Australia | Mars Society | Test habitats and space tools |

## IV. Earth-Bound Challenges

### A. Hazards of Space Debris

The prevalence of sub-orbital space debris of variable sizes around earth rises the potential danger to all space missions and vehicles. There are the circumstances where the spacecraft experiences collision with the debris and can lead to the Kessler Syndrome. (For example 1996, French satellite, 2009-Iridium Satellite). NASA Keeps tracking the debris and data of sub-orbital debris with the aid of the Department of Defense and Space Surveillance Network. Employing the data Joint Space Operation Center contribute to the interpretation of conjunction assessment to meet the human spaceflight criteria thereby reporting to the Johnson Spaceflight Center and Goddard Space Flight Center. Even though debris is carefully tracked, the threats come as a result of untraced debris of small sizes. For the manned mission, we need to launch and strand large seized orbital vehicles in LEO and have an increased chance of vulnerability to sub-orbital debris [23-25]. Nations and their quantity of debris are shown in Fig 4.

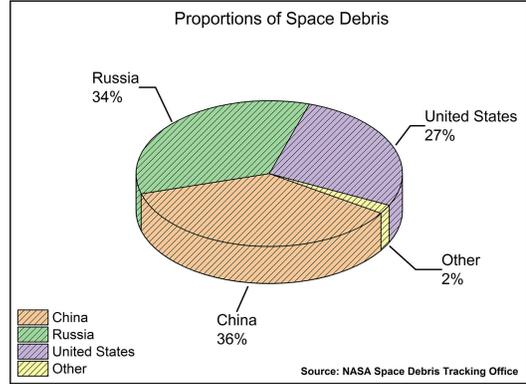

**Fig 4 Space Debris by Nations**

### B. Spaceship Rendezvous

In the scenario of the manned Mars mission, we need to launch an enormous number of spaceship segments into LEO to enable the assembly of Crew Transportation Vehicle. Launching a massive space vehicle aboard launchers is unsustainable due to the limitation of launch vehicle technology to assemble a transportation vehicle for Mars excursion. Hence the assembly requires refined orbital rendezvous and docking technology. We have successfully demonstrated this technology in the 1960s when Gemini 6A and Gemini 7 achieved a technological milestone. Since 1960 astronauts have gone through manual rendezvous (during Apollo and Shuttle Program) and automatic rendezvous (during ISS mission) by making use of Radar (Radio Detection and Ranging) and Lidar (Light Detection and Ranging) technology. And the Lidar has a better insight into future docking and proximity operations. Further, focusing on manual and autonomous rendezvous, both can be preferred in LEO (but autonomous is best). But in the case of Mars orbital rendezvous, autonomous proximity operations and docking is always preferred than manual method (considering the risk of crew safety and health fitness). Because a small misstep can lead to mission loss and space disaster. So, according to the statement of Dr. Dennehy (NASA GNC Technical Fellow) "Autonomous rendezvous and capture will be an integral element of going to Mars", it is absolute fact and we recommend to execute autonomous rendezvous and proximity operations in Mars orbit than manual [26-29].

### C. Orbital Refueling

Refueling of space vehicles and reusing of vented tanks or launcher components are the key technology of manned mission to drive the cost down. In refueling operations, significant criteria like fluid transfer, pressure control, pressurization, gauging, zero-boil off storage, mixing desertification, passive storage, and leak detection are to be considered for fuel management. The low gravity in space greatly influences the Deep Space Refueling process. Additionally, the technology of fluid transfer, liquid acquisition, and mass gauging have a low technology readiness level and need to be matured. So, we need to conduct more experiments and enhance refueling technology despite past experiments shown in table 2 [30-33].

**Table 2 – Experiments on Refueling and Management**

| Acronym | Experiment Name | Year |
|---------|-----------------|------|
| FARE | Fluid Acquisition and Resupply Experiment | 1992 |
| SFMD | Storable Fluid Management Device | 1992 |
| SHOOT | Super Fluid On-orbit Transfer | 1993 |
| TPCE | Tank Pressure Control Experiment | 1997 |
| VTRE | Vented Tank Resupply Experiment | 1996 |

### D. Recycle and Reusable Technology

Currently, we have extended our technology to reuse the launch vehicle components (For example Past space shuttle – first reusable launch system). To date, several space firms have demonstrated the reusable technology. Some of them were tabulated in table 3. Considering economic standard and low-cost space access, we need to extend and enhance the reuse and fabrication technology to space vehicles, fuel tanks, and degraded satellites (solar panels, batteries, and some reliable components).

**Table 3 – Reusable Lunching Systems**

| Launcher | Agency | Type | Status | Ref |
|---|---|---|---|---|
| New Shepard | Blue Origin | Sub-orbital | Under development | [34] |
| RLV-TD | ISRO | Sub-orbital | Successful flight test | [35] |
| Spaceship Two | Virgin Galactic | Sub-orbital | Three Successful flight test | [36] |
| Falcon 9 | SpaceX | Orbital | Operational | [37] |
| Falcon Heavy | SpaceX | Orbital | Operational | [38] |

### E. On-Orbit Construction and Assembly

On-Orbit constructions and assembly are one of the greatest challenges for future deep space exploration mission as well as Mars expedition. The manual on-orbit assembly has numerous threats and challenges like the effect of zero-gravity on physical health, exposure to solar irradiance, solar flare and eruptions, cosmic radiations, dynamics of astronauts, health, and energy aspects. Space firms like NASA, ROSCOSMOS, ESA are involved in developing artificial intelligence robots for on-orbit constructions to redress the above challenges. Hence for future missions (where large spaceships and space platforms are required), robotic on-orbit constructions assure 100% safe and secure assembly eliminating every on-orbit challenges [39, 40]. Levels of radiation exposure are shown in Fig 5.

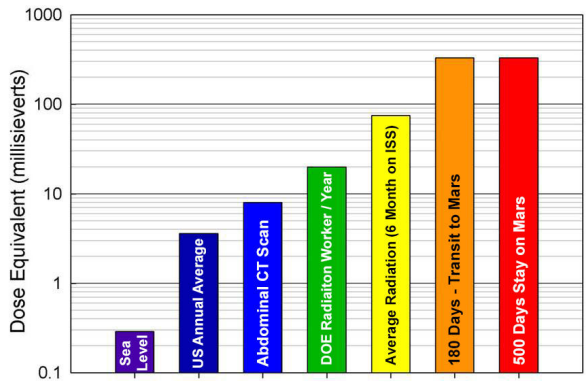
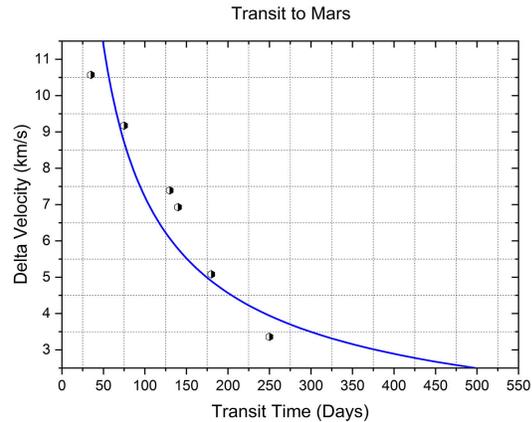

**Fig 5 Levels of Radiation Exposure on Earth & Mars (Image Courtesy: JPL/NASA)**     **Fig 6 Delta Velocity and Transit time to Mars**

### F. Achieving maximum delta-v (Δv)

Achieving maximum delta velocity (Δv) relies on possible orbit-raising maneuvers with the aid of chemical thrusters (thruster fairing on-off), gravity assist, and advanced propulsion system. Exploiting chemical propulsion systems, we can achieve a minimal delta velocity of 5.08 km/s (variable) and may take up to 180 days to transit from Earth to Mars. Longer transit time has increased exposure of crew to hazardous space environmental conditions [41]. However, exploiting the most preferred approach of NTR (Nuclear Thermal Rocket or Propulsion) proposed in many human mission architectures with maximum delta velocity of 8 km/s can minimize the transit time to 120-130 days approximately [42-43]. The transit time and delta-velocity relation are shown in Fig 6 and it shows the increase in delta velocity decreases the interplanetary transit time from Earth to Mars.

### G. Mars Trajectory

We have several trajectory options [44-47] and are classified into conjunction class and opposition class. Beneficial human class mission requires more scientific goals and its extent of accomplishment. So conjunction class trajectory is more favourable than the opposition class trajectory. Because conjunction class has Mars surface stay time of approx.350 days higher than the opposition class approx. 30 days and it minimizes the crew exposure to galactic cosmic radiation and solar flares being sheltered under the Martian environment. Opposition class trajectory has maximum exposure to cosmic hazard throughout Trans-Mars as well as Trans-Earth transits due to shorter surface stay period and additional requirement of Venus flyby (that makes closure vulnerability to the Sun and its hazardous elements). Opposition class mission increases the mission budget parallel to launch mass and propulsive energy, but in case of conjunction class mission – it follows the least energy path with minimal energy requirement (Hohmann's Transfer Trajectory). Hence, conjunction class mars mission is decidedly recommended for Human Mars Expedition and it is found to be the most proposed approach in various mission strategies. Comparison of the duration of crew exposed to various states of Mars expedition (both Conjunction and Opposition class trajectory) is shown in Fig 7.

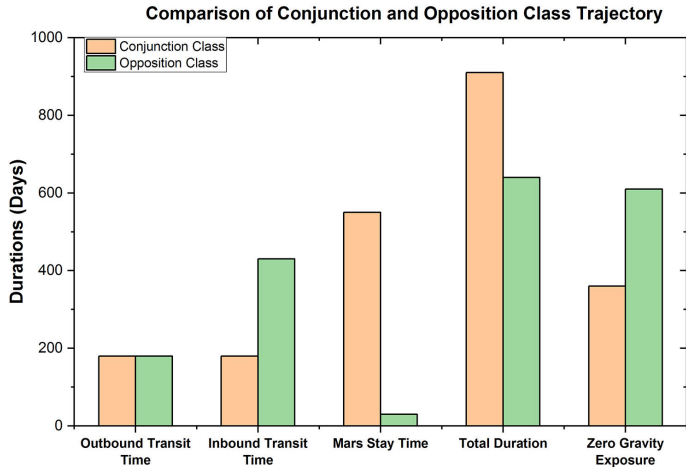

**Fig 7 Comparison of Trajectory Class**

## V. Interplanetary Challenges

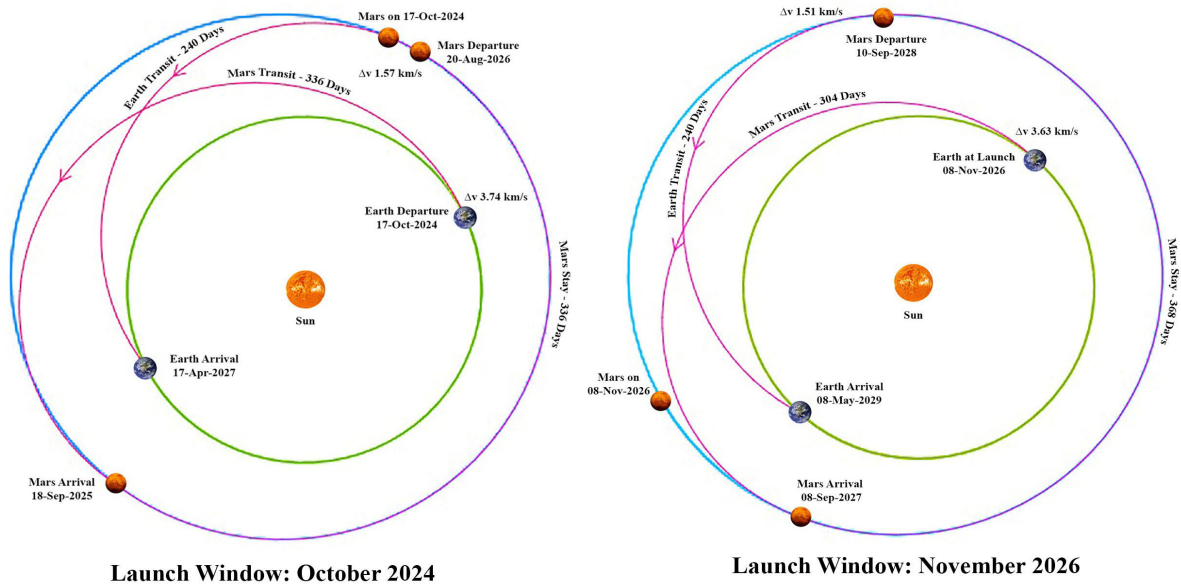

**Launch Window: October 2024**  **Launch Window: November 2026**

**Fig 8 Optimized Trajectory for Human Mars Mission between 2020 and 2040**

### A. Trajectory Option for Mars

Trajectory analysts have proposed numerous pathways for Human Mars Exploration [48-50]. Here, we have discussed the major two types of trajectory classes.

**Opposition Class:** Opposition class mission is often referred to as a short-stay mission where travelling astronauts spent most of their mission time in interplanetary space (both in outbound and inbound) with a surface stay of about 30-60 days. This class of trajectory is found risky and expensive because its optimized trajectory requires more Δv (higher energy transfer trajectory for transit back to earth) and longer duration in interplanetary space may subject to the exposure of galactic cosmic radiation. Additionally, the opposition class mission utilizes Venus flyby that highly enables the crew for closure exposure to Sun's hazardous elements. Further, the departure delta velocity of 7.0 km/s or above may decrease the success probability of Mars orbital capture and this requires maximized backward propulsion with high energy and fuel exhaustion [51].

**Conjunction Class:** Contradictory to opposition class, conjunction class is referred to as a long-stay mission where the crew spends most of their time on Mars surface (400-600 days) than stranding in interplanetary space. Because it makes the crew exposure to galactic cosmic radiation. But due to the benefit from the alignment of the planetary position of Earth and Mars, it grasps a minimal delta velocity of 3.36 km/s to follow a minimum energy path (Hohmann's Trajectory) for Mars vicinity thus affording the simplest way for Mars orbital capture upon Mars approach [51].

**Trajectory Assessment:** Several trajectory assessments showed that long-stay mission may expose the crew to harmful cosmic radiation [51]. But being shielded under the Martian environment is safer than spending much time in either Mars orbit or interplanetary space. The threat of radiation exposure on Mars can be minimized by the application of Mars Sub-Surface habitats or deep space habitats. We know that a human mission to Mars is no cheaper than a conventional mission, so the long term effort can be effectively exploited by performing various scientific observations and experiments during their long stay on Mars. Because short-stay mission may limit the long term experiments. Hence, we recommend conjunction class trajectory is best suited for a manned mission to Mars which is cheaper, safe, and effective in all aspects as compared to the opposition class. Moreover, in case of any emergency abort, the crew can vast-off Mars and follow high energy optimal path or free return trajectories to return back to earth [46]. Optimized trajectory for conjunction class with launch window at 2024 and 2026 is shown in Fig 8, the duration for the human-crewed mission from the 2022 launch window to 2037 is shown in Fig 9 [52] and delta velocity required for various abort options is shown in Fig 10.

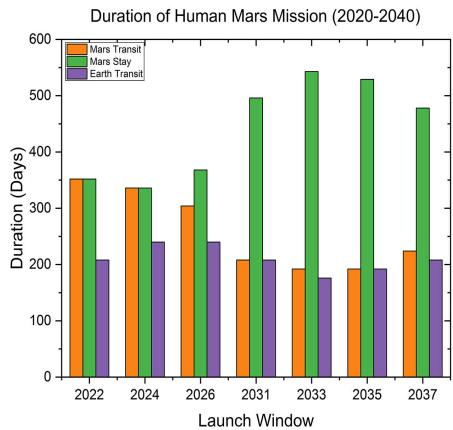 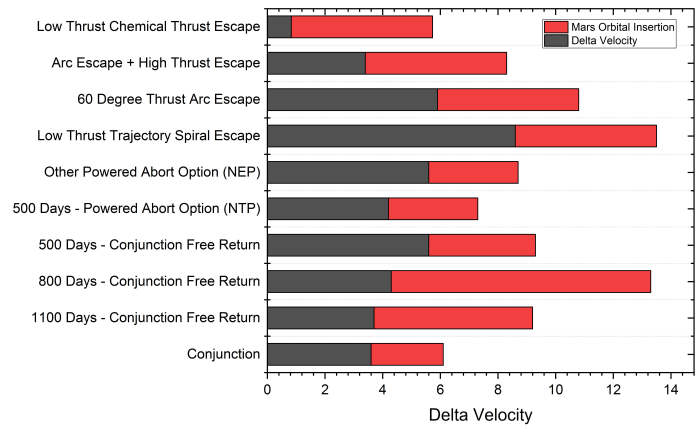

**Fig 9 Duration for Human Mars Mission (2020-2040)**    **Fig 10 Delta velocity requirements for abort options**

### B. Trajectory Correction Maneuver

Based on the study of failure analysis of conventional probes by M.K. Biswal [53,62], we have found that 1/4$^{th}$ of Mars probes encounters ignition engine issues caused as a result of the malfunction of the thermal control system. So, Mars Transit aboard a massive space vehicle via interplanetary coast may be subjected to low pressure, low temperature, and zero or microgravity environment directly affecting the zero-boil-off temperature of the cryogenic fuel and fuel pressurization systems. Therefore, improper fuel management and temperature imbalance may result in the inappropriate firing of thrusters plighted for mid-course or trajectory correction and maneuvering. Moreover, employing a larger delta velocity during the Earth departure stage may have serious concerns over trajectory correction and maneuvering of a massive space vehicle. Hence, proper fuel management and optimal delta velocity may curtail these challenges.

### C. Spaceship Management

Spaceship management and maintenance are some of the most challenging tasks for voyaging astronauts. Because of being under the microgravity and space radiation environment, the physical health of the astronauts may limit their access to the complete space vehicle and management. As we know that, the hardware and electronic equipment are the prime components advancing the space vehicle design and it gets degraded by the effect of long term exposure of harmful galactic cosmic radiation. This stands the most challenging quest for the crew aboard the spaceship. Hence, the construction of space vehicles with durable stuff (that are tested and validated in our ground-based laboratories with exposure to artificial radiation) and robust electronics are eminently endorsed. Further, it is desirable to employ artificial intelligence-based automated robots for complete management of the space vehicle to minimize the effort of the manual management system and manual detection of damaged or malfunctioned components [54].

### D. Effect of Radiation and Zero-Gravity

**Radiation:** Radiation and zero-gravity are the confronting challenges of human and interplanetary spaceflight. Astronauts have a greater threat of exposure to galactic and solar cosmic rays (GCR and SCR), and solar particle events (SPE). These are the natural phenomena spontaneously anticipating from our Milky Way galaxy and deep space and a threat to spaceflight safety and security systems. NASA has categorized the serious issues of GCR and SCR exposure into four human diseases. This includes carcinogenesis, cardiovascular disease as part of tissue degeneration, lifetime risks to the central nervous system, and acute radiation syndromes. So the health consciousness of the crew is very much significant for a successful manned mission [55, 56].

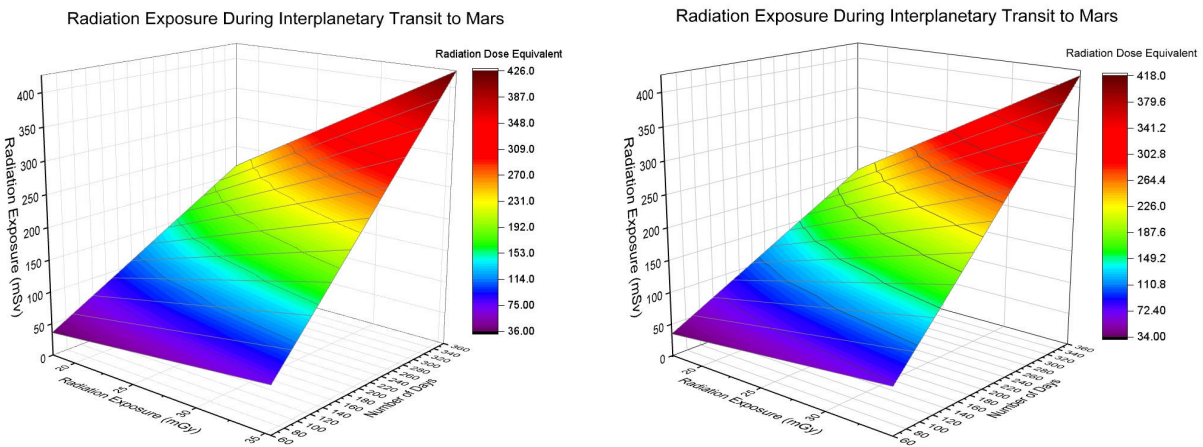

**Fig 11 and 12 Levels of Radiation Exposure during Interplanetary Transit to Mars**

In addition to this space, radiation can cause serious unrecoverable impairment of electronic components and hardware of space vehicles that might result in mission loss. Hence, sufficient thickness of radiation suit and sterling radiation shield is considered to avoid damage and degradation of the onboard circuitry system. It is because interplanetary spaceflight to Mars takes an average of 180-270 days manoeuvring the Hohmann's transfer trajectory may expose to harmful galactic cosmic radiation in the order of 1.16-1.18 millisieverts [57, 58] (extremely in opposition class trajectory due to additional requirements of Venus flyby). The level of radiation exposure to the number of days is shown in Fig 11 and 12.

**Zero-Gravity/Microgravity:** On a Human excursion to Mars, astronauts will experience three sorts of gravitational fields where one is during the interplanetary transit between planets, second is on the surface of Mars, and finally the third when they return back to earth. These three sorts of transition from distinct gravitational fields and zero-gravity can cause discord in brain coordination and functions, improper balance and orientation affecting the spatial orientation of brain, and motion sickness. NASA had performed various experiments aboard International Space Station (ISS) to understand the changes and impact of zero-gravity on the human body. The results showed that astronauts experience osteoporosis (bone density collapse due to the loss of bone minerals). Besides these concerns, inadequate ingestion of significant consumables and irregular exercise might result in loss of muscle strength and endurance [59, 60].

**Scientific Observation of Health caused by Zero-gravity:** An analysis from the Space Shuttle, Mir Program, and ISS Expedition mission showed that the crew experienced serious effects on their muscle system, bones, and cardiovascular activity [61]. After returning from the Space station crew had additional issue of improper blood pressure and blood circulation to the brain that displays the challenge in rehabilitation. But for the astronaut on a mission to Mars will remain in microgravity and zero-gravity environment over transit duration 180-270 days and may cause serious health issues. Hence sufficient countermeasures should be taken into consideration and the best way to confront this challenge is simulation or generation of artificial gravity on the spaceship. Further adequate food habitation and regular exercise will help make the astronauts remain fit and healthy throughout the mission.

### E. Solar Irradiance and Temperature

Solar irradiance plays a crucial role in power generation and temperature regulation during interplanetary transit to Mars. Lower solar irradiance would reduce the solar array output required for powering the spaceship and its operating devices onboard modules. It is because the availability of solar irradiance or intensity of sunlight steadily decreases as we move far away from the sun shown in Fig 13 [63]. In general assumption, the solar array output reduces from 3000 watts at 1366 W/m$^2$ intensity (at Earth) to 1000 watts at 588 W/m$^2$ (at Mars) and the record is assumed as per the solar cell configuration of NASA's Mars Reconnaissance Orbiter [64].

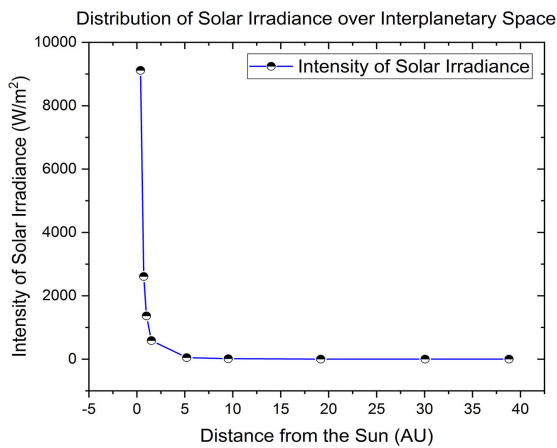

**Fig 13 Distribution of Solar Irradiance**

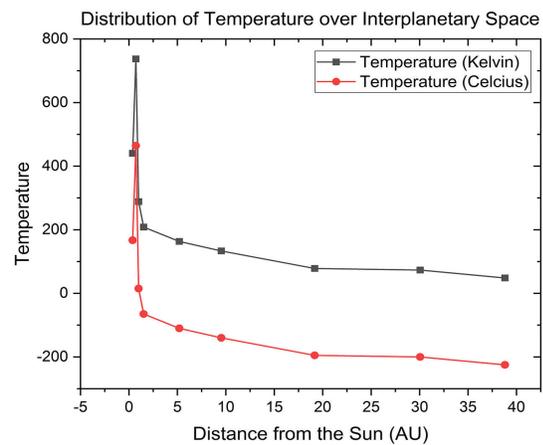

**Fig 14 Distribution of Temperature**

Corresponding to solar irradiance, the temperature of the interplanetary medium decreases by the function of inverse square law. We have shown the mean temperature variance of planets in Fig 14 based on the data provided by [65]. The temperature variance directly affects the thermal control systems of the space vehicle and proper thermal insulation is very much essential to keep astronaut warm and healthy in a hard environment. In addition to this, temperature variance also affects the fuel storage (zero boil-off and freezing), electronic components, and life support systems (for the growing plants or life aboard the spaceship).

### F. Effect of Nuclear Hazards

Nuclear energy is considered as the source of power for the future of deep space exploration and transportation systems. Mars scientists and engineers have proposed to manoeuvre nuclear electric propulsion and nuclear thermal propulsion rockets (NEP, NTR) to minimize the transit duration from Earth to Mars and to enhance space vehicle for a faster mission. Since the NEP and NTR are the emerging technologies for the advanced propulsion system and a progressive stride towards interstellar transit. Furthermore, it is the only choice that can meet our strategy of stellar and deep space exploration including interplanetary transportation systems. Because the potential for solar irradiance and solar power is limited beyond Mars orbit to employ solar electric propulsion system (SEP). Nevertheless, the NEP, NTR has direct effects on crew health affecting cardiovascular tissue, increasing the risk of cancer throughout the lifetime, and hereditary diseases. Hence, during the construction and assembly of the space vehicle, we endorse to mount the crewed module or base far away from the nuclear reactor with adequate shielding to overcome this challenge [66, 67].

### G. Isolation and Psychological Effects

A manned mission to Mars takes an average of 2.5 to 3 years that leads to the complete state of isolation/confinement of astronauts eventually affecting the behavioural and psychological patterns. Astronauts may encounter mood and cognition issue, the risk of anxiety, depression, digestive problems, loneliness, hypo or hypertension [68] as part of behavioural changes. Psychological changes include positive moods and relationships patterns with other crew members. A collaborative study performed by Russians and ESA's project members entitled "Mars-500" showed an increase in positive emotions among crew members [69]. But these results may remain undesirable because the actual environment of a space vehicle during interplanetary transit cannot be simulated on Earth. So, the experiences from Mars analog research stations and Arctic research stations can be treated for planning a mission to some extent.

The experiences from the crew of International Space Station (ISS) is inconsistent because the crew on transit to Mars encounter interrupted communication with their ground and family's relations than crew aboard ISS. But instead, they have to hold on for about 40 minutes for both transmitting and receiving a single message and hence this may lead to stressful situations and social concerns [70]. These changes are unanticipated and are far beyond how well trained and experienced they are. Therefore, it is substantial to sort out the astronauts who are physically and psychologically fit with multiple interdisciplinary skills. In addition to this, they must be capable of managing themselves during confined and stressful situations [71].

### H. Communication and Interplanetary Internet

**Communication:** Communication poses a crucial role in mission engagement and keeps the astronauts updated about the mission strategies from the ground. Identically it enhances the psychological personality of the crew and increases the success probability of mission accomplishment. The scientific demand for the human and robotic exploration missions to Mars and beyond is expanding, so high bandwidth and uninterrupted advanced communication relays are highly required. Since at a distance of 1.5 to 2.5 AU (Mars encounter and beyond) the communication interlink is limited to 24 to 40 minutes and it may not be an efficient approach to stay tethered with operating science mission orbiters and crewed space vehicles [72,73]. Because human mission beyond Moon and LEO is completely new and the crew is exposed to the inexperienced environment. So it is very significant to keep the crew updated about the safety measures and their next move. Additionally, the technological unavailability of persistent communication coverage for manned vehicles (other space probes in Mars orbit or on the surface), proper interlink during superior solar conjunction, the need for simultaneous control over multiple proximity operations on Mars vehicles (Spaceships, orbiters, landers, and rovers), and the exigency to access the Deep Space Network (DSN) for current mission trends are the some of the significant challenges on the way to communication systems [74]. Detailed Review on Advanced Communication Technologies for Human Mars Exploration and their countermeasures are explained in [75].

**Interplanetary Internet:** Interplanetary Internet provides easy access to reliable scientific resources for the crew and enables the public to get updated with live coverage from interplanetary space. This mode of consistent internet and communication from the public directly to the crew may minimize the mental stress (due to good appreciation and encouragement from public interactions) thereby increasing their interdisciplinary activities. Further, interplanetary internet enables the onboard crew to perform various scientific studies or research works as part of their free time with massive access to research papers (optional assignment). The current trend of networking technology can able to accomplish an interlink transfer of ~100-2015 Mbps for downlink and ~10-25 Mbps for uplink via Earth-Mars trunk line [74] and need to be upgraded for advanced communication relay. In addition to this, NASA is currently planning either to park two-three telecommunication relay orbiters in HMO or MSO to form an integrated constellation architecture for providing increased bandwidth and data transfer rate or to park in Earth-Sun Lagrangian point for deep space communication access with optical fibre and laser-guided communication system [72, 76].

### I. Cryogenic Fuel Management during Interplanetary Transit

Alike on-orbit or in-space refueling discussed in section 4, cryogenic fuel management (CFM) is necessary during the extended presence of humans on an exploration mission to Mars, and beyond. Because propellant plays a vital role in transiting space vehicles from a point to the destination and assists in performing sturdy mid-course and trajectory correction maneuver. Nevertheless, fuel depletion may prompt in the decline of the mission and it is necessary to effectuate proper CFM procedure. Further, the future exploration mission greatly relies on three factors of the extraterrestrial based cryogenic fuel management system, they are management of cryogenic propellant (liquid hydrogen, liquid methane, liquid oxygen), storage, and distribution. As we discussed in section 4, active thermal control, demonstration of cryogenic propellant management, fuel transfer, instrumentation leak detection, liquid acquisition devices, low-gravity mass gauging, passive thermal control, pressure control, refrigeration, storage, system feed testing, and zero boil-off are the key technology for fuel management [30, 77].

These elements of CFM pose a substantial challenge for fuel management on the way to interplanetary transit to Mars. Efforts to address this challenges were developed in NASA's ground-based laboratories include Cryogenic Propellant Operations Demonstrator (CPOD), Experimentation for the Maturation of Deep Space Cryogenic Refueling Technology (MDSCR), In-Space Cryogenic Propellant Depot (ISCPD), Zero Boil-Off Tank (ZBOT) [77,78]. Till date, accomplished CFM technology and the demand for future were clearly reviewed by Chato [77]. We have shown the approximate time duration required for propellant storage and technology readiness level for CFM in Fig 15 and Fig 16.

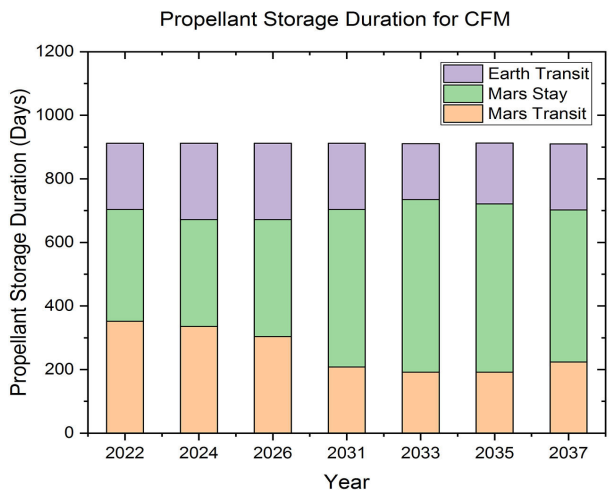
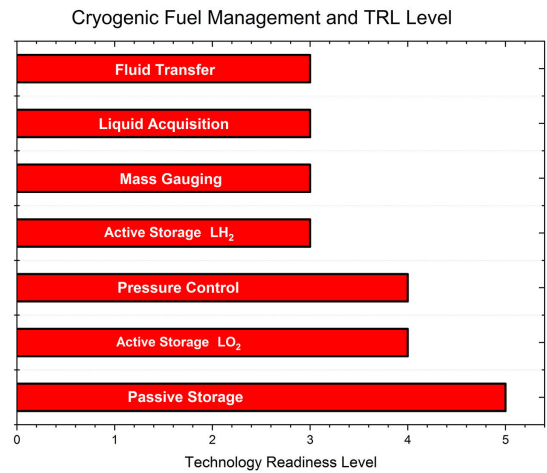

**Fig 15 Duration required for CFM**  **Fig 16 Technology Readiness Level for CFM**

### J. Waste Recycling and Management

Long duration transit to Mars my result in the generation of large amount to trash that ultimately plunges the crew to biological and physical hazard. The entire trash generated includes exhaled $CO_2$ gases from astronauts, used waters, human wastes, biological wastes, solid wastes, nuclear wastes, and medical wastes. Amidst these, biological, nuclear and medical waste poses a serious threat to crew and may increase the probability of crew sickness and a prolonged threat to cancer. So, the disposal of these trashes requires adequate attention and disposal procedure (anaerobic digester) with consideration of crew safety and risk [79]. According to Lockhart, a crew of four can generate a waste of about 2.5 tons per year. In consequence manned mission to Mars for 2-3 years may result in the generation of 7.5-8 tons [80]. Hence, the measure for recycling and reusing is an ideal approach for a sustainable interplanetary transit rather than conventional trash disposal and venting, because the cargo re-supply and availability is limited from the Earth. The current trend of recycling method followed aboard International Space Station can be further modified and enhanced to accommodate adequate waste management, water recycling, and oxygen generation during interplanetary transit to Mars [81,82].

### K. Extra-Vehicular Activity

Space colonization highly necessitates the potential capabilities of hours of extravehicular activity performed by the astronaut outside their shielded spacecraft with sophisticated skills and mature technology. EVA plays a great role in accomplishing space assembly, servicing, and space vehicle management and can be extended to provide ecstatic owing to the psychological effects experienced by the astronauts during interplanetary travel. EVA requires a bounded environment in terms of pressure, temperature, and the concentration of oxygen with the additional capability of supplying water and food, temperature regulation, pressure retention, and waste collection within the suit [83].

But the challenges that affect EVA are direct exposure of astronauts to the galactic cosmic radiation, elements of solar flares, difficulty in balancing their momentum in zero gravity, mechanical hazards and the risk of losing in space as a result of a defective tether. In addition to this galactic particles or micro asteroids travelling with a relative velocity of 10 km/s may hit and puncture the space suit pushing into a critical situation [84]. Further, an analysis on these challenges by Pate-Cornell [85] grouped the overall EVA challenges into eight categories namely failure of airlock and life support system, fire in the suit, mechanical and radiation accident, separation, spacesuit failure, and de novo events. Hence, astronaut equipped with good EMU (Extravehicular Maneuvering Unit) suit may reduce the risk of losing in space. Moreover, healthy space suit with high tolerance (when exposed to vacuum environment) and multiple capabilities can increase the robustness and longevity of EVA. The suit should be developed concerning the experiences encountered by EVA astronauts [86] aboard ISS and other human spaceflight program. Furthermore, NASA's Johnson Space Center is currently involved in the development of Robonaut (Robotic astronaut's assistant) that can enhance the efficiency of EVA hours with improved capabilities of strength, dexterity, and mobility [87].

### L. Mars Approach and Orbit Capture

Once we vast-off from the Earth, the next destination is approaching Mars. Accurate targeting and successful Mars Orbital Insertion (MOI) thereby diminishing the delta velocity is very significant for a successful mission. Because, once the space vehicle remains ineffective in capturing orbit, it is very difficult to maneuver back the spaceship to its destination point. As we discussed earlier, the factors such as the delta velocity, fuel management, proper functioning of navigation, maneuvering, and ignition system determines the success probability of Mars Orbital Insertion. However, many inexperienced robotic spacecraft like ESA's Mars Express and ISRO's Mars Orbiter Mission (MOM) has successfully demonstrated excellent Mars Orbital Capture in their first attempt [88-90]. But, in the case of the crewed mission, it is extremely recommended that experiences and lesson learned from past successful Mars probes can be doubtlessly applied for orbital capture. Since inaccurate targeting, timing, and malfunction of breaking engine could pose the space vehicle stranded in interplanetary space or could cause the vehicle destroyed in Mars atmosphere and mission tragedy [91].

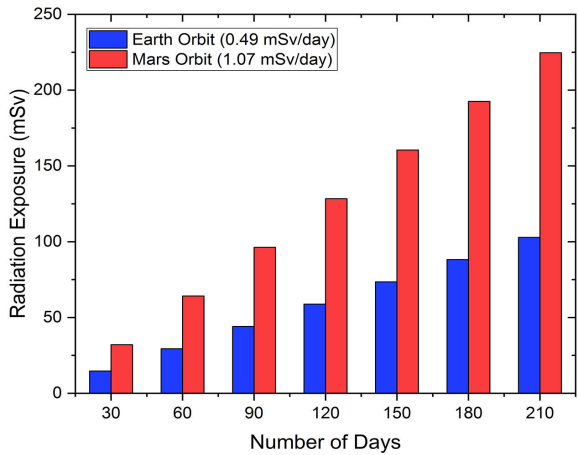
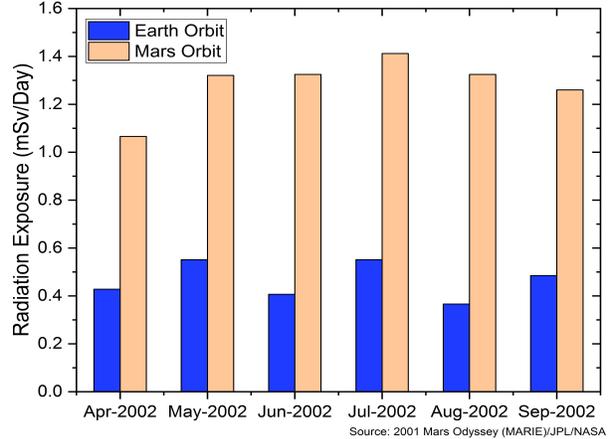

**Fig 17 Radiation Exposure at Earth & Mars**     **Fig 18 Comparison of Radiation Dose (Earth & Mars)**

## VI. Mars-Bound Challenges

### A. Hazards of Exposure to Cosmic Radiation

Consequent to successful Mars Orbital Insertion, the Mars-bound challenges commence, as the crewed space vehicle need to strand in Mars orbit until further instructions from the ground for mission inception. Because the spaceship undergoes preliminary checks and validation of its components and communication relay system. So, during their stay in Mars orbit, the astronauts along with their spacecraft are exposed to high dosage of harmful galactic and extragalactic cosmic radiation as compared to low earth orbit. This radiation levels range from a minimum of 1.07 millisieverts per day to a maximum of 1.4 millisieverts per day (measured by MARIE experiment aboard 2001 Mars Odyssey) ultimately increasing the possibility of prolonged cancer and related diseases [92-93]. The challenges and the consequences of radiation exposure were explicitly elucidated in section IV and additional complication was explained in [94-96]. Hence, astronaut being sheltered under the Martian environment (on the surface of Mars) seems safer than stranding in low Mars orbit (LMO). Furthermore, we have graphically presented the comparison of radiation levels at both Earth and Mars orbit in Fig 17 and the amount of radiation assimilated by the astronauts depending on the duration of stay in LMO in Fig 18.

### B. Hazards of Asteroid Impact

Astronauts aboard space vehicle in orbit or spacecraft orbiting the red planet have vulnerable to the hazard of an asteroid impact and damaging of spacecraft components. The asteroids sizes vary from micro to macro asteroids. These asteroids are ejected from the main asteroid belt due to the probabilistic collisional events occurring at a distance ranging from 2.4 to 3.4 AU from the Sun with a relative velocity of 5.4 to 8.0 km/sec [97]. Larger asteroids can be mitigated or destroyed by directing into the Mars atmosphere, but the problem is with micrometeoroid or micro-asteroid that are travelling at a relative velocity of >10km/sec can cause severe damage to the spacecraft component: it includes rupture of spacecraft fuel tank through impact and penetration, affecting the spacesuit of an astronaut during extravehicular activity, and depleting the solar arrays and affecting the power production. However, this challenges cannot be completely eradicated, but proper attention required while fabricating the sensitive component of space vehicles (i.e., fuel tank, hardness of solar array, and glasses of life-support systems). Before initiating the interplanetary exploration missions sufficient thickness of walls of the fuel tank is considered during fabrication and before stepping out for extra-vehicular activity. The challenge of asteroid impact will continue to exist for future interplanetary missions beyond Mars as the main asteroid belt lies between Mars and Jupiter and spread up to hundreds of kilometers (approximately 150 million kilometers) in interplanetary space with asteroid size ranges from 50 to 150 km in diameter [98-99].

## C. Communication and Solar Power Production

**Communication:** We have already discussed the challenges of the interplanetary communication system in section 5. However, human exploration missions predominantly require advanced communication systems to guide and land the spacecraft modules safely on Mars. Because a small misstep during EDL phases my cost mission tragedy as two-way communication interlinks span about 20 to 45 minutes and unreachable throughout solar conjunctions. Hence enhanced and advanced communication system (i.e., laser-guided communication system that has been successfully demonstrated by NASA [100]) either via Mars Telecommunication Orbiter (MTO) or real-time decisiveness is ideally recommended for crewed missions [101].

**Solar Power:** Power production at Mars appears to be the most strenuous task for Martian satellites since the intensity of solar irradiance evanesces from Earth to Mars shown in the Fig 13. Hence for a crewed missions thousands of watts are essentially required. So larger solar arrays capable of outstretching their solar cells are preferred to meet the energy requirements to power the space vehicle and estimated power production rate is 100 watts per square meter (at a solar intensity of 588 W/m$^2$). Further, this power option is limited during the Mars solar conjunctions when Earth and Mars are far from each other having Sun at the median point for a period of 10-15 days [102]. Therefore, we can alternatively exploit radioisotope thermoelectric generators (RTG) to afford the basic power necessities thereby mounting the RTG at a safe distance from the crewed module with proper shielding (to avoid the effects of nuclear radiation) [103]. Data transfer rate and mean power generation capacity of some Mars spacecrafts are shown in table 4.

**Table 4 Power Generation and Data Transfer rate of operating spacecraft at Mars**

| Power Productions at Mars Orbit [104-108] | | | Data Transfer Rates from Earth to Mars [109-110] | |
|---|---|---|---|---|
| Spacecraft | Area of Solar Array | Power Generation | Up-link Transfer Rate | Downlink Transfer Rate |
| Mars Express | 11.0 m$^2$ | 650 watts | 2-128 kbps | 28 kbps |
| 2001 Mars Odyssey | 7.00 m$^2$ | 750 watts | 8-256 kbps | 8 kbps |
| Mangalyaan | 7.56 m$^2$ | 840 watts | 5-40 kbps | 8 kbps |
| MGS | 6.65 m$^2$ | 980 watts | 8-128 kbps | 8 kbps |
| MAVEN | 12.0 m$^2$ | 1135 watts | 500 kbps | 3-4 mbps |
| ExoMars TGO | 17.5 m$^2$ | 2000 watts | 8-26 kbps | 26 kbps |
| MRO | 20.0 m$^2$ | 2000 watts | 500 kbps | 3-4 mbps |
| **Mean** | **12.0 m$^2$** | **1194 watts** | MAVEN has highest achieved data transfer and download rate than other space probes i.e., **500 kbps to 4 mbps** | |
| Approximate Power Rate = **100 watts/square. meter** | | | | |

## D. Planetary Clearance and Mars Atmospheric Entry

Mars Entry, Descent, and Landing is the sturdy assignment for both crewed and planetary landers due to the limitation of uncertainty in predicting the natural hindrance (prevalence of the variable environmental condition and dust storms) and EDL technology. Despite this challenge, Mars entry may disrupt the communication system and can cause damage to the landing module. Therefore, this obstacle can be eluded by reporting the crews in advance (grasped in Mars orbit) about the environmental condition forecasted by Mars relay orbiters. It ensures crew about the planetary clearance and their next move. Usually, this allows the crewed landers to perform orbital entry instead of direct entry which is considered as the safest approach and recommended for crewed landing [111]. The path for both direct and orbital entry is shown in Fig 19. Because it reduces the entry velocity for increased ballistic coefficient and faster aerocapture, the risk of crash landing due to limited EDL period, and design flexibility. Further, it enables the crews and their landing modules for an effective preparation to plunge into the Mars atmosphere.

In addition to this, the factors such as the geometry of aeroshell, diameter of parachute, entry velocity and ballistic coefficient pose a design and technical challenge for Mars landers. From the first landing attempt to the current state of technology, the geometry and mass factor of landers is limited to the diameter of the aeroshell (4.5m) and gross landing mass up to 0.9-ton. So we need to expand the diameter of aeroshell as well as parachute. Nevertheless, larger aeroshell requires larger payload fairing that seems to be more expensive than conventional launch vehicles. Alternatively, the aeroshell diameter can be enhanced with the use of hypersonic inflatable aerodynamic decelerator (HIAD) to cut down the terminal velocity and ballistic coefficient thereby increasing the drag force that will drastically reduce the terminal velocity for dynamic landing. We have not reviewed much about Mars EDL challenges as it was technically reported by R.D. Braun [7]**.**

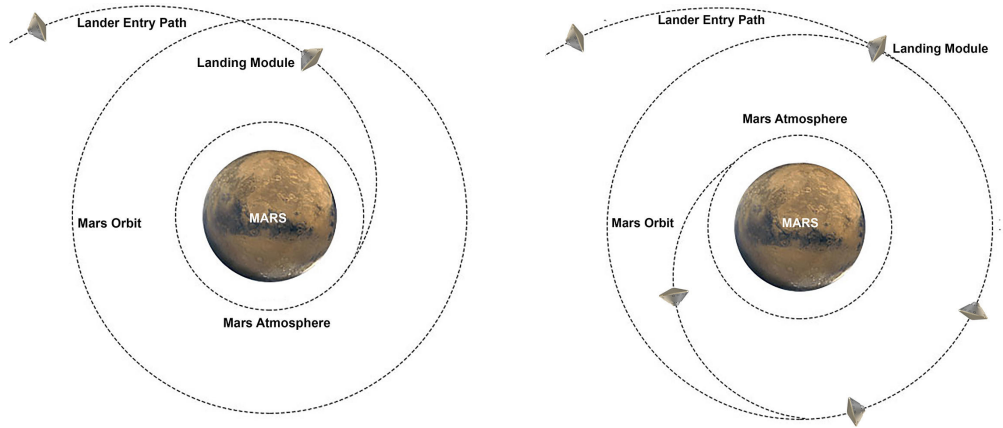

**Fig 19 Trajectory Path for Direct and Orbital Mars Entry**

## VII. Planetary Surface Challenges

### A. Scientific Landing Site / Exploration Zone

Exploration zone with good scientific interest and resource determines the success and sustainability of the crewed mission on Mars. The scientific site should meet all necessities for the crews and should have affordable native resources for exploitation to keep alive the crew during extended surface stay mission. NASA has identified forty-seven candidate landing site for robotic and manned exploration [112-113]. Of these Meridiani Planum seems to be the best site for first Crewed Mars landing and Base establishment. Because, it holds an ideal site for promising resources which includes the potential for water extraction, raw materials for infrastructure purposes, and minerals. Meridiani Planum is located at 50°N and 50°S with an elevation of below +2 km (MOLA). Additionally, it enables crews for practicing planetary cropping and plantation, food production with efficient solar power production as it lies near-equatorial latitude, and facilitates for accomplishing multidisciplinary scientific goals in terms of atmosphere, astrobiology and geosciences. Features and scientific interest of candidate landing site (Meridiani Planum) were completely reviewed by Clarke [114].

### B. Planetary Environment

**Density of Mars Atmosphere:** Due to the thin atmosphere of Mars, the planet is incapable of shielding its surface from being exposed to harmful cosmic radiation and pose a threat to the living astronaut on the surface. Similarly, its lean atmosphere with lower density forbids lander modules from faster aerocapture thereby limiting the EDL period [111]. In addition to this, the composition of Mars atmosphere $CO_2$ (95%) and $O_2$ (0.17%) stands a challenge for the astronaut to breathe outside their spacesuit [115].

**Low-Gravity Environment:** Astronauts on Mars gets exposed to the low gravity environment affecting the periodic pattern of heartbeat, rate of blood flow, weakening the muscle and bone density, and physical movements. The human body takes time to adapt their internal organs to sustain their presence under low gravity environment. Hence, these issues can be managed by frequent practicing of physiotherapy and physical exercises [116-118] or simulating artificial gravity on Mars.

**Solar Irradiance and Power:** The challenges of solar power do exist at every extremity beyond LEO. For a manned mission, this complication comes during the interplanetary voyage, stranded in Mars orbit, and on Mars surface. However, the intensity of solar irradiance weakens from orbit to the surface and the mean power production rate is about 20 watts per square meter (Source: InSight Mars Lander) [119]. Hence, the structure of extendable solar arrays can be employed for mass electricity production but instead, the nuclear thermoelectric generators (NTG) will be the ideal choice for power source on the surface (during the day and night). Furthermore, the intensity of solar irradiance influences the environmental temperature that poses a challenge against manned Mars exploration to stabilize the thermal stability [120] and this challenge can be addressed with the application of Mars sub-surface habitats.

**Temperature and Pressure:** The frequency of temperature on Mars varies for every Martian year. Findings and observations from diverse spacecraft have shown that the temperature variance ranges from the lowest 120K near poles to the highest of 293K at the equator with an average of -210K [121,122]. Hence at this range of temperature, the crew may experience complication in maintaining the thermal stability of both habitats and their body's internal temperature to keep them warm against hypothermia [123]. Contradictory to temperature the pressure varies from 400 Pa to 870 Pa according to the seasonal pattern. So, this challenge can be addressed by deploying Mars sub-surface habitat [124] to balance the thermal stability during the day and night. Simulated graph of air density, solar flux, temperature, and pressure distribution on Mars shown in Fig 20 [125].

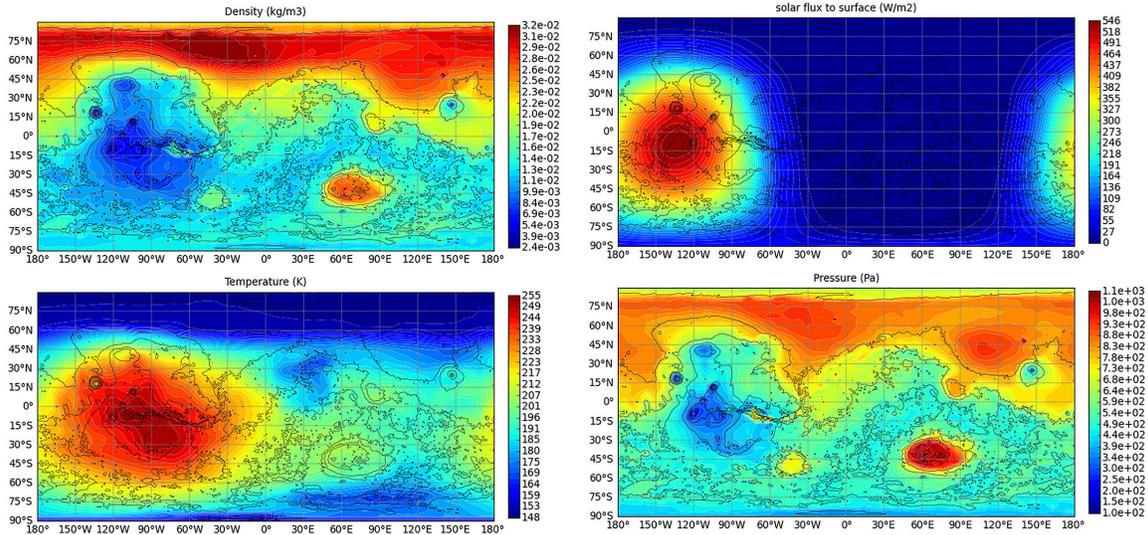

**Fig 20 Simulated graph of air density, solar flux, temperature, and pressure distribution on Mars (as of Jan 2021). Image Courtesy: Mars Climate Database**

### C. Exploitation of Resources

The challenge associated with the exploitation of resources is locating a robust site for exploration as well as extraction. Because the distinct site is associated with divergent distribution and concentration of resources, the form at which they exist, and the quantity of contaminants from the aspect of planetary protection. Since transportation of resources from different sites to the base is limited due to the constraints in surface mobility and lack of long-range rovers. Further, the unavailability of the testbed to demonstrate and validate ISRU instruments under a critical and low gravity environment poses a technical challenge on the surface. Furthermore, the uncertainty in system reliability and integration remains the greatest challenge at the very beginning of the Mars Base foundation [126, 127]. Current NASA plans for the future Mars In-Situ Resource Utilization is briefly presented by G. Sanders [127].

### D. Base Construction and Surface Mobilization

For a limited crew member at the initial stage of colonization, the habitats and the other modules can be exported from the Earth. But for a larger number of the population, a Mars Base is required and construction of this massive base using labour force is not obvious due to the vulnerability of Martian Environment and exhaustion of limited survival resources. Hence robotic based construction is beneficial from the perspective of crew health and also in retaining survival resources [128]. Similarly, system reliability and its extended operation in a critical environment remain inconsistency due to technical challenges such as unproven technologies in the appropriate testbed and solar power deficiency. In addition to this, base construction is supported by load transportation systems from various resource mining sites. This mode of transportation may prompt the systematic servicing and repairing of vehicles [129, 130]. Artist concept of Mars Base is shown in Fig 21.

### E. Communication Interlink from Earth

The communication challenge is discussed earlier in the interplanetary challenge section. But it is significant for the crew on Mars to stay tethered and updated about the mission plans. The communication interlink is interrupted during the superior solar conjunction for every synodic period. Hence this challenge can be addressed either by parking communication relay satellites on high non-kaplerian orbit to stride uninterrupted communication [131] or Mars Telecommunication Orbiters (MTO) into Mars orbit prior to the mission expedition.

### F. Ethical Challenge and its Implications

Astronauts either during their interplanetary transit or extended period of mission may encounter the serious effect of ethical challenge. Despite NASA's policy, crew members aboard spacecraft may likely to experience the risk of in-space pregnancy due to their mental and extreme stressful situation throughout the mission. Besides this ethical issue can put other crew members in risk due to additional requirement of survival resources, intensive care, and attention required for the new born astronaut and mother. The complexity of space environment and microgravity may cause serious health implications to mother and child and can lead to fatal state [132]. It is still uncertain that how far this ethical challenge can be prohibited, but the prime cause for this challenge (i.e. mental stress and psychological health) can be improved to some extent with live interaction and conversation with either family members or publics on the Earth via interplanetary internet. Further, it is recommended to select the astronauts with multidisciplinary skilled and psychologically fit in terms of social abilities.

### G. Space Policy and Human Civilization

Once the progression for human civilization commences, crews from various nations may encounter the political challenge concerning the conflict of interest between nations. Because human civilization nation's appropriation of Martian land violates the space policy of 1967 "United Nations Treaty on Principles Governing the Activities of States in the Exploration and Use of Outer Space, Including Moon and other Celestial Bodies". According to that treaty or policy, the space and the astronomical bodies over Solar system are free for research related activities to all nations without any differentiation or discrimination [133]. But none of the astronomical body or planetary space cannot be expropriated by any nation. Therefore, we recommend that having a good mutual understanding and universal conflict of interest between countries will help make the civilization towards peaceful planetary exploration and permanent base establishment. Further, it we are planning for a return or round-tip, we may encounter the hazard of back contamination from the red planet as part of Planetary Protection Policy [134-135]. And this challenge can be remitted with proper screening and medical checks aboard International Space Station before landing the crews on Earth.

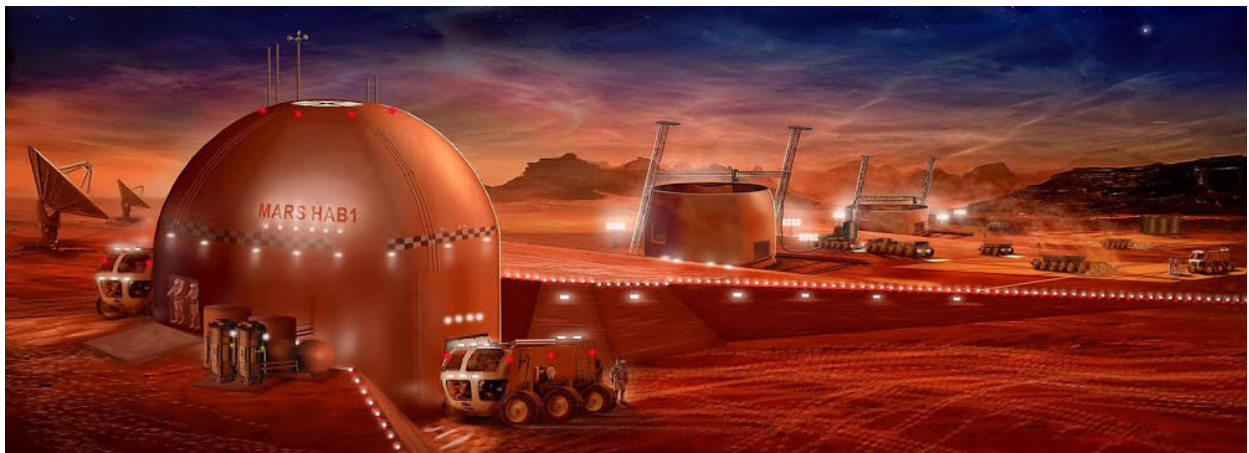

**Fig 20 Artist concept of Mars Base Establishment and Mars Base Construction
(Image Courtesy: James Vaughan - NASA's 3D-Printed Habitat Challenge)**

## VIII. Conclusions

Mars is the only planet in our solar system to harbor terrestrial life as the planet is enriched with all resources required for human civilization. The leap for a human-crewed mission to Mars as well as outer planets is likely to fulfil in next decades as the aerospace progression has achieved a milestone in exploring the outer solar system planets through robotic and unmanned spacecrafts. Further, Mars is the next destination for establishing human civilization beyond low earth orbit, and a platform to demonstrate all of our advanced technology. Human interplanetary mission to Mars may encounter diverse challenges with the complicated environment and it is necessary to overcome these challenges for prosperous and sustainable missions. Henceforth, we have delineated every possible challenge en-route to Mars. The challenges were briefly discussed under the categorization of terrestrial, interplanetary, and planetary challenges. Subsequent to these challenges, some possible recommendations were reviewed in their appropriate section as countermeasures. Our study reports the complete state of challenges except for Entry, Descent, and Landing challenges which was technically described by R.D. Braun. Surprisingly, an analysis reveals that the interplanetary space beyond the extremity of Earth and Mars have a promising environment to sustain the reliability of complex hardware necessitate for human Mars missions [136]. Finally, we expect that our study may provide a useful framework for the mission planners and space entrepreneurs to consider and mitigate these challenges from the perspective of a successful mission.

## Acknowledgements

I (Malaya Kumar Biswal) would like to thank to my research guide **Dr. Ramesh Naidu Annavarapu., Ph. D** (author of this paper) and my lovable friends for their financial assistance for the conference participation. Also I would like to extend my sincere gratitude to all the department professors for their guide and innovative vision to pursue my research career. Further, I would like to acknowledge the Department of Physics for providing me great opportunity to pursue my research and advance my career.

## Dedication

I (Malaya Kumar Biswal) would like to dedicate this work to my beloved mother late. **Mrs. Malathi Biswal** for her motivational speech and emotional support throughout my life.

## Conflict of Interest

The authors have no conflict of interest to report.